\documentstyle[12pt]{article}
\advance \textheight by \topskip
\begin{document}
\begin{titlepage}
\hbox to \hsize{\hfil }
\hbox to \hsize{\hfil {\bf hep-ph/9701201}}
\hbox to \hsize{\hfil January, 1997}
\vfill
\large \bf
\begin{center}
Universality of the preasymptotics
in the hadron and photon diffraction
\end{center}
\vskip 1cm
\normalsize
\begin{center}
{\bf S. M. Troshin and N. E. Tyurin}\\[1ex]
{\small  \it Institute for High Energy Physics,\\
Protvino, Moscow Region, 142284 Russia}
\end{center}
\vskip 1.5cm
\begin{abstract}
We note that it is premature to make
a conclusion on the multiplicity of the Pomerons
on the basis of the available experimental data
since the interactions have a preasymptotic nature.\\[2ex]
PACS number(s): 12.40.Pp, 13.60.Hb, 13.85.Lg
\end{abstract}
\vfill
\end{titlepage}
The straitforward interpretation of the recent HERA data on the
deep--inelastic scattering together with the analysis of the
 data on hadron--hadron
scattering in  terms of the Regge model could lead to the
unexpected conclusion
on the existence of the  various Pomerons \cite{land} or
the  various manifestations
of the unique Pomeron
in the different processes depending on the typical
scale of the process \cite{brt}.
 The approaches \cite{dl,pr} contending the
dominance of the soft Pomeron do not rule out
 existence of the
hard Pomeron either.

Indeed, soft hadronic reactions imply that the Pomeron's intercept
$\alpha_{\cal{P}}=1.08$ \cite{land}, small--$x$ dependence of the
structure function $F_2(x,Q^2)$ leads to
$\alpha_{\cal{P}}=1.4$-$1.5$ \cite{h1,zeus} and the measurements of
the diffractive cross--section in the deep inelastic scattering
provide $\alpha_{\cal{P}}=1.23$ \cite{dfr}.  So, does this mean
that  we have few Pomerons or we have few different manifestations
of  the same Pomeron depending on the particular process?  Probably
both options are not to be considered as the firm ones, since the
experimental data  used for these statements were obtained at  not
high enough energies where, in fact, the preasymptotic regime of
interactions does take place.  The above conclusions are based on
the presumed dominance of the Pomeron contribution already in the
preasymptotic energy region and do not take into account the
unitarity effects which become very essential as one goes beyond
 this  region. What is called a Pomeron  is to be interpreted as a
true asymptotical contribution of the driving  mechanism.

 In this note we argue that all the three classes of the processes
described above are related to the similar mechanisms and the
corresponding energy dependence of the cross-sections can be well
described by the universal functional energy dependence  of the
type $a+b\sqrt{s}$. Such dependence is valid for the preasymptotic
energy region only and beyond this region unitarity changes the
 picture drastically.  We   consider  for  illustration  the
unitarized chiral quark model \cite{us94}. However, the conclusions
seem to be of a more general meaning.

 In this model the elastic scattering amplitude in the impact
parameter representation has the following form \begin{equation}
F(s,b)=U(s,b)[1-iU(s,b)]^{-1} \label{um} \end{equation} where
$U(s,b)$ is the generalized reaction matrix which in the case of a
pure imaginary amplitude is \begin{equation}
U(s,b)=ig(N-1)^N[1+\alpha\sqrt{s}/\langle m_Q\rangle]^N
\exp(-\tilde Mb).\label{uf} \end{equation} In Eq. (\ref{um}) $g>
0$, $N$ is the total number of the constituent quarks in the
colliding hadrons, $\tilde M=\sum_{Q=1}^N m_Q/\xi$ and $b$ is the
impact parameter of the colliding hadrons \cite{us94}.  Fit to the
total $hp$ cross-sections  gives small values for the parameters
 $g$ and $\alpha$ ( $g,\alpha\ll 1$) \cite{pras}.  It means that at
$s\ll s_0$ the second term in the square brackets in Eqs.
(\ref{um}) and (\ref{uf}) is small and we can expand over it.  The
numerical value of $s_0$ is determined by the equation $|U(s,0)|=1$
and is \cite{pras} \[\sqrt{s_0}=2\,\mbox {TeV}.\] The value of
$s_0$ is on the verge of the preasymptotic energy region, i.e.  the
Tevatron energy is at the beginning of the road to the asymptotics.
Evidently the HERA energy range $W(=\sqrt{s_{\gamma p}})\leq 300$
GeV is in a preasymptotic domain.

The above model gives the linear with $\sqrt{s}$ dependence for the
total cross--sections according to Eqs. (\ref{um}) and (\ref{uf}):
\begin{equation} \sigma_{tot}^{hp,\gamma p}=a+b\sqrt{s},\label{lin}
\end{equation} where parameters $a$ and $b$ are different for
different processes.  It was shown \cite{pras} that this dependence
is in a good agreement with the experimental data.

The same dependence for the total cross--section of $\gamma^* p$
scattering is implied by the small--$x$ behavior of the structure
function $F_2(x,Q^2)$ observed experimentally \cite{h1,zeus} and
obtained  in the model \cite{eps}:  \begin{equation}
F_2(x,Q^2)=a(Q^2)+b(Q^2)/\sqrt{x}.\label{f2} \end{equation} The
experimental data  indicate the critical behavior of the function
$b(Q^2)$ at $Q^2\simeq 1$ (GeV/c$)^2$.

The third value for the Pomeron intercept $\alpha_{\cal{P}}=1.23$
has been obtained from the analysis of the experimental data on the
diffractive cross--section in deep--inelastic scattering \cite{dfr}
 where the dependence of $d\sigma^{diff}_{\gamma^* p\to XN}/dM^2_X$
on $W$  was parametrized according to the Regge model and the
Pomeron dominance has been assumed:  \begin{equation}
d\sigma^{diff}_{\gamma^* p\to XN}/dM^2_X\propto
(W^2)^{2\alpha_{\cal{P}}-2}.\label{rd} \end{equation}

The data  demontrate  linear rise of the differential
cross--section $d\sigma^{diff}_{\gamma^* p\to XN}/dM^2_X$ with $W$,
i.e.  we observe here just the same functional dependence on the
c.m.s. energy as it was observed for $\sigma_{tot}^{hp,\gamma
p,\gamma^* p}$.  Regarding the preasymptotic nature of the
interaction mode we arrive  to the universal c.m.s. energy
dependence in the framework of the model used above.

Indeed, in the framework of this model the hadron inelastic
diffractive cross--section is given by the following expression
\cite{usdf}:  \begin{equation} \frac{d\sigma^{diff}_{hp\to
XN}}{dM_X^2}\simeq \frac{8\pi g^*\xi^2}{M_X^2}\eta (s,0),\label{ds}
\end{equation} where \[ \eta (s,b)=\mbox{Im} U(s,b)/[1-iU(s,b)]^2
\] is the inelastic overlap function.

At the preasymptotic energies $s\ll s_0$ the energy dependence of
inelastic diffractive cross--section resulting from Eq. (\ref{uf})
 is determined by the generic form \begin{equation}
\frac{d\sigma^{diff}_{hp\to XN}}{dM_X^2} \propto
a+b\sqrt{s}.\label{ds1} \end{equation} Inelastic diffractive
cross--section for the $\gamma^* p$ interactions can be obtained
using for example VMD model, i.e.  \begin{equation}
\frac{d\sigma^{diff}_{\gamma^* p\to XN}}{dM_X^2} \propto
a(Q^2)+b(Q^2)W.\label{ds2} \end{equation} The same functional
dependence can be obtained using the "aligned jet" model \cite{alj}
along with the unitarized chiral quark model \cite{us94}.

It should be noted here that the above linear dependences for the
cross--sections of different processes is the generic feature
associated with the preasymptotic nature of the interaction
dynamics at $s\ll s_0$. As one goes above this  energy range the
function $|U(s,b)|$ is rising and when $|U(s,0)|\geq 1$ the
unitarity starts to play the major role and provides the $\ln^2 s$
rise of the total cross--sections at $s\gg s_0$ \cite{us94} and
also the following behavior of the structure function $F_2(x,Q^2)$
\begin{equation} F_2(x,Q^2)\propto\ln^2( 1/{x}) \end{equation} at
$x\rightarrow 0$ \cite{eps}.  At the same time unitarity leads to
the decreasing dependence of the inelastic diffractive
cross--section at $s\to\infty$ \begin{equation}
\frac{d\sigma^{diff}}{dM_X^2} \propto
\left(\frac{1}{\sqrt{s}}\right)^N.\label{ds3} \end{equation} for
the  $hp$, $\gamma p$ and $\gamma^* p$ processes \cite{usdf}.  Such
behavior is associated with the antishadow scattering mode which
develops  at small impact parameters at $s> s_0$.

Thus, we might expect the different asymptotic and universal
preasymptotic behaviors for the different classes of the
diffraction processes.

To summarize, we would like to emphasize that the unified
description of the processes of $hp$, $\gamma p$ and $\gamma^* p$
diffraction scattering with the universal cross-section dependence
on the c.m.s. interaction energy is possible.  For the illustration
we used the unitarized chiral quark model which has a
nonperturbative origin and leads to the linear c.m.s. energy
dependence of the cross--sections  in the preasymptotic energy
region for the above processes. Universality of such preasymptotic
 behavior agrees with the  experiment.

The assumption on the existence of the different Pomerons results
from the use of the asymptotic formulas in the preasymptotic energy
region and the neglect of the unitarity effects at higher energies
beyond this preasymptotic region. It should be taken with certain
caution.

 \end{document}